\documentclass[superscriptaddress,preprintnumbers,amsmath,amssymb,nofootinbib, prd, twocolumn]{revtex4}
\usepackage{amssymb}
\usepackage{amsmath}
\usepackage{hyperref}
\usepackage{color}
\usepackage{graphicx}
\newcommand{\be}{\begin{equation}}
\newcommand{\ee}{\end{equation}}
\newcommand{\bea}{\begin{eqnarray}}
\newcommand{\eea}{\end{eqnarray}}
\newcommand{\tr}{{\rm Tr}}
\newcommand{\Eqref}{Eq.~\eqref}

\begin{document}

\title{Towards phase transitions between discrete and continuum quantum spacetime from the Renormalization Group}
\author{Astrid Eichhorn}
\affiliation{\mbox{\it Perimeter Institute for Theoretical Physics, 31 Caroline Street N, Waterloo, Ontario, N2L 2Y5, Canada}
\mbox{\it E-mail: {aeichhorn@perimeterinstitute.ca}}}

\author{Tim Koslowski}
\affiliation{\mbox{\it Department of Mathematics and Statistics, University of New Brunswick, Fredericton, New Brunswick E3B 5A3, Canada}
\mbox{\it E-mail: {t.a.koslowski@gmail.com}}}

\begin{abstract}
We establish the functional Renormalization Group as an exploratory tool to investigate a possible phase transition between a pre-geometric discrete phase and a geometric continuum phase in quantum gravity. In this paper, based on the analysis of \cite{Eichhorn:2013isa}, we study three new aspects of the double-scaling limit of matrix models as Renormalization Group fixed points: Firstly, we investigate multicritical fixed points, which are associated with quantum gravity coupled to conformal matter. Secondly, we discuss an approximation that reduces the scheme dependence of our results as well as computational effort while giving good numerical results. This is a consequence of the approximation being a solution to the unitary Ward-identity associated to the U(N) symmetry of the hermitian matrix model. Thirdly, we discuss a scenario that relates the double scaling limit to fixed points of continuum quantum gravity.
\end{abstract}

\maketitle
\section{Introduction}
The functional Renormalization Group (FRG) is an exploratory tool in the investigation of quantum gravity models. In \cite{Eichhorn:2013isa} we showed how the FRG can be used to recover the double scaling limit of matrix models for 2D Euclidean quantum gravity as a Renormalization Group (RG) fixed point. The purpose of this paper is to further develop the use of the FRG as a tool in the investigation of matrix models of 2D Euclidean quantum gravity. These models are the prototype of discrete models of quantum gravity such as Euclidean and Causal Dynamical Triangulations  \cite{Ambjorn:1991pq, Ambjorn:2012jv,Ambjorn:2011cg}, tensor models  \cite{Rivasseau:2011hm, Rivasseau:2012yp}, and group field theories \cite{Boulatov:1992vp,Freidel:2005qe,Oriti:2007qd,Oriti:2011jm}.  The ultimate goal of this research is to provide a qualitatively well-understood and quantitatively precise analytical tool to study the continuum limit in four-dimensional models of discrete quantum spacetime.

The particular results of this paper are the following:
\begin{enumerate}
 \item Besides the double scaling limit there exists a series of multicritical points, which correspond to the continuum limit of pure quantum gravity coupled to conformal matter. We discover these multicritical points in sec.~\ref{sec:RGSym} as fixed points of the FRG with good matching of the critical exponents and dimensionless ratios of coupling constants. 
 \item It turns out that the tadpole approximation to the beta functions does not only simplify computations, but also leads to scheme independent results. Further, this approximation solves the tadpole approximation to the unitary Ward-identity. This leads to an improvement of the numerical results for critical exponents compared to \cite{Eichhorn:2013isa}, where we found a 50 \% discrepancy of the critical exponent for the pure gravity model compared to the known analytic value. This quantitative improvement arises, as the tadpole approximation is a good approximation for fixed points with small critical values of the couplings. Besides, it provides a self-consistent implementation of unitary symmetry.
 \item The continuum limit of matrix models is well understood in terms of a lattice interpretation of Feynman graphs of a matrix model, as a limit in which the lattice constant approaches zero. This straightforward geometric picture is not available in the FRG description. Nevertheless, there is a link between the FRG approach to matrix models and continuum quantum gravity (see sec.~\ref{connection}).
\end{enumerate}
Let us now revisit the foundations of the approach of \cite{Eichhorn:2013isa} before introducing their implementation in sec.~\ref{FRG}.
\subsection{Renormalization Group and double-scaling limit}
The idea behind the matrix- (resp. tensor-) model approach to quantum gravity is to express the quantum gravity path integral as a sum over discrete triangulations or more generally tesselations. This discrete sum can be translated into matrix or tensor models \cite{Weingarten:1982mg}. In two-dimensional quantum gravity, based on the Einstein-Hilbert action \footnote{Note that the restriction to the Einstein-Hilbert action is somewhat arbitrary at the microscopic level, in particular from the FRG perspective. In fact, it could well be that the existence of a continuum description of quantum spacetime requires higher-order operators.} and a summation over  topologies, the corresponding matrix model is of the form
\be
Z= \int \mathcal{D}\varphi\, e^{N\left(- \frac{1}{2} \tr \varphi^2 + \frac{g_4}{4} \tr \varphi^4 \right)},
\ee
where $\varphi$ is an $N \times N$ hermitian matrix, for reviews see, e.g., \cite{AlvarezGaume:1991rm,Ginsparg:1991bi,David:1992jw, Di Francesco:1993nw,Ambjorn:1994yv}. The continuum limit is obtained when $N \rightarrow \infty$ (i.e., infinitely many degrees of freedom contribute), and $g_4 \rightarrow g_{4\, c}$. If these two limits are taken separately, only spherical topologies contribute to the partition function.
To include all topologies, one should observe that the partition function admits an expansion in Feynman graph topologies
\be
Z= \sum_h Z_h\, N^{2(1-h)},\label{topsum}
\ee
where $h$ is the number of handles. It is possible to keep contributions of all topologies, if the $\frac{1}{N}$ suppression of higher topologies is compensated. This is possible,  since  $Z_h \approx (g_{4\,c}-g_4)^{(1-h)(2-\gamma_{st})}$ as $g_4 \rightarrow g_{4\,c}$. Thus, if one takes the double scaling limit \cite{Douglas:1989ve,Brezin:1990rb,Gross:1989vs}
\be
(g_4 - g_{4\, c})^{\frac{2- \gamma_{\rm st}}{2}} N = C,\label{doublescaling}
\ee
where $C$ is a constant, while $N \rightarrow \infty$ and $g_4 \rightarrow g_{4\, c}$, then all $Z_h$ contribute to the large $N$-limit of $Z$. To understand the connection to the Renormalization Group, we realize that the double-scaling limit dictates a particular scaling of $g_4$ with $N$.
We can write \Eqref{doublescaling} as
\be
g_4(N) = g_{4\, c} + \left( \frac{N}{C}\right)^{-\frac{2}{2- \gamma_{\rm st}}}.
\ee
This is the structure of a solution to the linearized RG flow in the vicinity of a fixed point\footnote{Given a beta function $\beta_g = \mu \partial_{\mu}g (\mu)$, we can linearize it around a fixed point at $g_{\ast}$: $\frac{\partial \beta_g}{\partial g}\Big|_{g = g_{\ast}} (g- g_{\ast})=0$. This is solved by $g(\mu) = g_{\ast} + c\, \left(\frac{\mu}{\mu_0} \right)^{- \theta}$, where $\theta= - \frac{\partial \beta_{g}}{\partial g} \Big|_{g= g_{\ast}}$, and $c$ is a constant of integration and $\mu_0$ a reference scale.}. Thus the double scaling limit corresponds to an interacting fixed point of the Renormalization Group of the matrix model, in which $N$ plays the role of an RG scale.
The critical exponent $\theta = \frac{4}{5}$ is related to the exponent $\gamma_{\rm st}$ in the usual notion for these models by $\theta = \frac{2}{2- \gamma_{\rm st}}$, i.e., $\gamma_{\rm st} = -\frac{1}{2}$.

Note that in the two-dimensional matrix-model case the sum \eqref{topsum} is not summable, so the partition function $Z$ exists only as a formal sum. In particular, while $g_{4\, c}$ is the radius of convergence for the perturbative expansion of the partition function at fixed topology, $g_{4\,c}$ plays no such role in the sum over all topologies.
Accessing the double scaling limit as an RG fixed point does not provide a novel way to perform this sum. Instead, the FRG framework provides a way to determine whether there exists a consistent scaling that relates $g_4$ and $N$, such that a contribution of all topologies to the formal sum can be retained in the large $N$-limit. 
The FRG allows us to \emph{derive a consistent scaling} for the double-scaling limit.
Whether or not such a scaling exists, and what the value of the critical exponent is, is independent of the question whether the partition function converges. In particular, the RG will give meaningful results for the scaling exponent of the double-scaling limit \emph{even in cases where the expression for the partition function does not converge}.
In higher dimensions, there are indications that the partition function is summable in the double-scaling limit \cite{Bonzom:2014oua, Dartois:2013sra,Kaminski:2013maa} and contributions from higher orders in the $1/N$ expansion \cite{Gurau:2010ba} can be retained consistently. 
The FRG allows us to access tensor models corresponding to $d=4$ dimensions, where other methods that work successfully in the matrix-model case, break down. The FRG will thus provide a method to derive the scaling exponent(s) of the double-scaling limit. 
Further, the FRG can also be applied to models with a matrix Laplacian, which are asymptotically free \cite{Geloun:2012bz}. In these models, the FRG could be of considerable use to study the strongly-interacting ``infrared" limit, where a phase transition could lead to a ``condensed" phase of discrete building blocks, see, e.g., \cite{Oriti:2013jga}.

In this paper, we will further establish the Functional Renormalization Group as a useful tool to study matrix models, providing a starting point for research on $d=4$ dimensional tensor models.
\subsection{Renormalization Group scale in matrix models}
Let us now expand on the use of $N$ as an RG scale, as first proposed in \cite{Brezin:1992yc}, see also \cite{Ayala:1993fj,Higuchi:1993np,Bonnet:1998ei}.
Applying Renormalization Group tools in quantum gravity could 
seem futile: The RG sorts quantum fluctuations according to a scale, separating large-scale from small-scale fluctuations, and integrating them out according to this organizing principle. In quantum gravity, where all possible geometries
 are included in the path-integral, no fixed notion of scale exists. Every configuration 
comes with its own notion of scale.
It thus seems that the use of RG techniques in quantum gravity requires to break the diffeomorphism invariance (the symmetry that ensures background independence) of gravity by singling out one field configuration to 
organize all other quantum fluctuations into large-scale or small-scale ones w.r.t. this preferred field configuration. Hence, it seems that background independence, a crucial 
requirement of quantum gravity, is incompatible with RG techniques.

Different answers have been found to this apparent problem: In continuum formulations, the background field method can be used in combination with diffeomorphism Ward-identities to introduce a background and fluctuation field \cite{Abbott:1980hw, Reuter:1996cp,Donkin:2012ud}, in such a way that background independence is ensured. In these approaches, the topology and dimensionality of quantum configurations are fixed, and this background structure is sufficient to then set up a useful background-field approach. 
Within Causal Dynamical Triangulations, proposals for the RG flow were recently advanced \cite{Ambjorn:2014gsa,Cooperman:2014owa}, based on equipping each fundamental building block with a length scale $a$, and then proceeding similarly to lattice field theory.

In completely background-independent approaches, where not even fiducial background structure is allowed, topology is not fixed, and geometry is emergent, the notion of scale can clearly not be related to a momentum-scale as in standard quantum field theories. The only possible notion of scale is inherent to the model, which ``knows" nothing about momentum or length scales. The notion of scale is related to the number of degrees of freedom that have been 
integrated in the path integral. Then, a distinction of UV and IR is possible: The model-inherent scale is infrared, if most degrees of freedom have been integrated out, and ultraviolet, if most degrees of freedom are not yet integrated out. It is then obvious that the matrix size $N$ is the only possible notion of scale in pure matrix/tensor models. \\
\subsection{Multicritical matrix models and conformal matter}
In an extension of our earlier work \cite{Eichhorn:2013isa}, we will discuss models beyond the pure gravity case here, which correspond to conformally coupled matter-gravity theories. The motivation to study these is clear: Our universe contains both matter and gravitational degrees of freedom, which are coupled. Thus matter degrees of freedom are important in the gravitational RG flow and vice-versa, see, e.g., \cite{Dona:2013qba} for the continuum case. It is thus highly interesting to study matrix models which correspond to tesselations of surfaces including dynamical matter. A model for these theories was discovered in \cite{Kazakov:1989bc}, where multicritical points for matrix models were found.
If we generalize the matrix model to allow for a potential of the form
\be
V(\phi) = \frac{1}{2} \tr \phi^2 - \frac{g_4}{4} \tr \phi^4 - \frac{g_6}{6} \tr \phi^6+...,
\ee
then this still corresponds to a model of random surfaces, with the tesselations including squares, hexagons, octagons and so on. If we allow these couplings to change sign, then some configurations will come with a negative weight, already suggesting that this model contains more than just gravitational degrees of freedom. E.g. consider a typical Feynman graph, which is a triangulation of a Riemann surface and consider a ``contamination'' with squares. The additional squares can be viewed as a soldering of two triangles and this soldering can be physically interpreted as a hard dimer \cite{Staudacher:1989fy}.
In fact, there exists a tower of  multicritical models, which correspond to two-dimensional gravity coupled to conformal matter.
They are specified by two integers $(p,q)$, which determine the central charge $c= 1- 6(p-q)^2/(p q)$. In the case of multicritical matrix models, it turns out that $(p,q) = (2, 2m-1)$, with $m=2,3,...$. One can then evaluate the critical exponent $\gamma_{\rm st}$ in Liouville theory, and obtains the result $\gamma_{\rm st} = -m+3/2$.
For the matrix models, the following pattern of critical points emerges: Setting $g_{n}=0$ for $n>n_{\rm max}$ and letting the $g_{n-1}, g_{n-2}$ etc. alternate in sign, with $g_4<0$, leads to different universality classes in the continuum limit. For the $m$th such multicritical point, $n_{\rm max} = 2m$, which is characterized by $m-1$ positive critical exponents in the RG approach \cite{Brezin:1992yc}. The smallest of these takes the value $\gamma_{\rm st}=\frac{3}{2}-m$, in agreement with the continuum result. (Note that different conventions for the definition of $\gamma$ are sometimes used in the literature. We follow \cite{AlvarezGaume:1991rm}.) Translated to the critical exponent at an RG fixed point, this implies $\theta_m = \frac{4}{2m+1}$.
The existence of further relevant directions \cite{Brezin:1992yc}, i.e., parameters that require tuning  to reach the phase transition in the large $N$-limit, is suggestive of the existence of further degrees of freedom. 
One can easily conjecture that these additional degrees of freedom are matter \cite{Kazakov:1989bc}.
As a new test of our Renormalization Group method, we will search for the fixed points corresponding to the double-scaling limit at these multicritical points, see, e.g., \cite{Ambjorn:1992gw}, and compare the results for the critical exponents with the exact values.
\section{Functional Renormalization group for matrix models}\label{FRG}
The Wetterich equation \cite{Wetterich:1993yh} is a functional differential equation for the effective average action $\Gamma_k$ of a quantum field theory, which contains the effect of quantum fluctuations at momenta $p^2>k^2$, and encodes the effective dynamics of low-energy effective fields. For general reviews of the method, see \cite{Berges:2000ew,Gies:2006wv}.
In \cite{Eichhorn:2013isa} we adapted this equation to the setting of matrix models, where no momentum scale exists. We thus introduce an infrared cutoff scale $N$, and write an infrared regulator $R_N(a,b)$ as a function of the matrix indices $a,b$ and the cutoff scale $N$, such that
\bea
\underset{a/N\rightarrow0, b/N\rightarrow0} {\rm lim} R_N(a,b)_{ab\, cd}  &>&0,\label{IRsup}\\
\underset{N/a\rightarrow0, N/b \rightarrow 0}{\rm lim} R_N(a,b)_{ab\, cd} &=&0\label{IRlim}\\
\underset{N\rightarrow \Lambda \rightarrow \infty} {\rm lim}R_N(a,b)_{ab\, cd} &\rightarrow&\infty.\label{UVlim}
\eea
Since the quadratic term $\tr(\phi^2)$ is dimensionless, i.e., it scales as $N^0$, we used  a dimensionless regulator of the form 
\be
R_N(a,b)= \left(\frac{2N}{a+b}-1 \right) \theta \left(1- \frac{a+b}{2N} \right),\label{equ:dimlessReg}
\ee
in \cite{Eichhorn:2013isa},
that is modeled after Litim's optimized cutoff for the continuum case \cite{Litim:2001up,Litim:2000ci}.

Including the IR-suppression term $\frac{1}{2}\tr(\phi\,R_N\,\phi)= \Delta S_N$ into the path integral
\be
\mathcal{Z}_N = \int_{\Lambda} d \varphi\, e^{-S[\varphi] - \Delta S_N[\varphi]+J \cdot \varphi},
\ee
 then allows us to define the effective average action by a modified Legendre transform.
\be
\Gamma_N[\phi] = \underset{J}{\rm{sup}} \left( J \phi - \ln \mathcal{Z}_N\right)- \Delta S_N[\phi]. 
\ee
The scale dependence of the effective average action is then described by
\be
\partial_t \Gamma_N = \frac{1}{2} {\textrm{tr}} \left(\Gamma_N^{(2)}+ R_N \right)^{-1} \partial_t R_N,\label{Wetteq}
\ee
where $t = \ln N$. Herein $\Gamma_N^{(2)} = \frac{\partial^2}{ \partial \phi_{ab} \partial \phi_{cd}} \Gamma_N$.
\subsection{Symmetric theory space and truncation}
To derive $\beta$ functions from \Eqref{Wetteq}, we write the effective action as a sum of local operators multiplied by scale dependent couplings. 
Our model has a $U(N) \times Z_2$ symmetry, such that all operators of the form $\tr\left(\phi^{i_1}\right)\dots \tr\left(\phi^{i_n}\right)$ with $i_1+\dots +i_n$ even are generated and should be included in the effective action. The space of action functionals that are linear combinations of these operators is the $Z_2 \times U(N)$-symmetric theory space. One now has to find a way to truncate this space to a finite subspace that can be dealt with in practice without throwing out those operators that are relevant for the physical system we aim to describe.
An organizing principle in this theory space is provided by the scaling dimensionality. 
In our case, where no notion of momentum scales exist, the scaling dimensionality is related to the matrix size $N$: The requirement of a well-defined large-$N$ limit of the matrix model allows us to derive a consistent (albeit not unique) canonical scaling of couplings:
For single-trace couplings $\bar{g}_i$ of operators $\tr\, \phi^i$ and their dimensionaless version $g_i$ we obtained in \cite{Eichhorn:2013isa}, see also \cite{Douglas:1989ve}
 \be
g_i= \frac{\bar{g}_i N^{\frac{i-2}{2}}}{Z_{\phi}^{i/2}},\label{canondim}
\ee
where $Z_{\phi}$ is a wave-function renormalization, occurring as the prefactor of the quadratic term in the potential.
Similarly we have that
\be
g_{i_1\dots i_n}= \frac{\bar{g}_{i_1\dots i_n} N^{\frac{i_1+\dots+i_n}{2}+(n-2)}}{Z_{\phi}^{\frac{i_1+\dots+i_n}{2}}},
\ee
for multitrace couplings $g_{i_1\dots i_n} \tr \left(\phi^{i_1}\right) \dots \tr \left(\phi^{i_n}\right)$.
Here we have taken into account that couplings corresponding to multitrace-operators have a lower canonical dimensionality, to account for the additional traces.

As only couplings with positive dimensionality are relevant and correspond to free parameters, we conclude that the theory has no free parameters at the Gau\ss{}ian fixed point. At an interacting fixed point, such as that corresponding to the double-scaling limit, new operators can be shifted into relevance as interactions modify the scaling dimensions. The canonical dimensionality nevertheless provides a useful organizing principle: If we assume that the contribution of quantum fluctuations to the scaling dimensionality, $\eta_i$, is bounded, this provides a guiding principle to set up a truncation: It should first include single-trace operators up to a certain number of fields, and then double- and triple-trace operators up to a given canonical dimensionality. Note that similar considerations have been backed up by explicit evaluations of scaling dimensions in continuum quantum gravity \cite{Falls:2013bv}.
We will use this reasoning to define useful truncations.

\subsection{Extended theory space with symmetry breaking operators}
An IR-suppression term that divides the matrix entries $\phi_{ab}$ into IR and UV degrees of freedom necessarily breaks the $U(N)$ symmetry of the matrix model. To construct a regulator that depends on matrix indices, one needs to introduce at least a constant matrix $X$ with components
\begin{equation}
 X_{ab}=a\,\delta_{ab},
\end{equation}
which allows one to introduce the operator  $\left(\Delta\,\phi\right)_{ab}=(a+b)\,\phi_{ab}$ as
\begin{equation}
 \Delta\,\phi:=X\,\phi+\phi\,X.
\end{equation}
This operator is closely related to the 2-dimensional Grosse-Wulkenhaar Laplacian $\left(\Delta\,\phi\right)_{ab}=(a+b+1)\,\phi_{ab}$, which can be used to set up an FRG approach to noncommutative scalar field theory \cite{Sfondrini:2010zm}. Due to this analogy, we will use the term ``matrix Laplacian'' for this operator.

As a  consequence of introducing the regulator, the RHS of the flow equation will also contain operators with insertions of the constant matrix $X$, i.e., operators of the form $\tr\left(\phi^{n_1}\,X^{m_1}\,...\right)...\tr\left(...\right)$ (with $\sum_i n_i$ even). In other words, the RG flow has to be set up in the extended theory space, which is the space of linear combinations of these more general operators, because the $U(N)$-symmetric theory space, which is a subspace of this full theory space, is not left invariant by the flow.

The dimensionality of the operators with additional  insertions of $X$ can be inferred in two ways: 
From inspection of \Eqref{equ:dimlessReg}, we realize that, as ${\rm tr} X \sim N$, this sets a dimensionality of 1.
Alternatively, one can consider
the geometric picture that underlies the 2-dimensional Grosse-Wulkenhaar model \footnote{Note that the analogy with the Grosse-Wulkenhaar model can only be made for the geometric structures, but not for $\phi$, which is dimensionless when treated as a 2D scalar field, whereas it is important to the matrix model approach to quantum gravity that $\phi$ possesses dimension $N^{\frac 1 2}$.}. There, the matrix size $N$ is related to the noncommutative scale $\theta$, which has dimension $l^{-2}$, which is the same dimension as the Grosse-Wulkenhaar Laplacian $\Delta$. Again, one concludes that $X$ should carry one unit of dimension in $N$.

This dimensionality of $X$ (resp. $\Delta$) has the useful effect of dimensionally suppressing operators with an insertion of $X$ compared to the $U(N)$ symmetric operators that are obtained by excising the $X$ insertions. Thus, one expects that the operators with $X$ insertions will be irrelevant.\footnote{Once one realizes that the flow equation does not preserve the $U(N)$-symmetric theory space, it becomes a natural question to ask what happens if one uses a standard regulator built from the matrix Laplacian $\Delta$, e.g., a regulator of the Litim form
\begin{equation}\label{equ:LitimReg}
 R_N = \left(N-\Delta\right)\theta\left(1-\frac{\Delta}{N}\right).
\end{equation}
In the explicit calculations that we performed for this paper, it turned out that the flow generated using the dimensionful regulator (\ref{equ:LitimReg}) reproduces all the qualitative features of flow generated using the dimensionless regulator (\ref{equ:dimlessReg}), and that the quantitative differences where small. It may at first seem surprising that the dimensionality of the regulator does not influence the qualitative features of the flow, but it is simply a consequence of the fact that we investigated the flow of a pure matrix model under the change of matrix-size $N$ and that both regulators are ``small matrix''-suppression terms.}
\subsection{Projection on a truncation}
The practical use of the flow equation involves truncations of the theory space. Unfortunately, the RHS of the flow equation will in general not be of the form of the truncation. Thus one needs to find a prescription to project the RHS of the flow equation onto the monomials in the truncation. In \cite{Eichhorn:2013isa}, we concerned ourselves only with the symmetric theory space and where then able to discern the monomials in the truncation by inserting field configurations of the form $\phi=v\,X$, after which we where able to perform the operator traces as simple sums. It is obvious that this simple projection rule is not sufficient for the full theory space. We thus refine the projection rule. In a first reading, the following considerations can be skipped, as the approximation that we will introduce in sec.~\ref{sec:tadpole} and the results in sec.~\ref{sec:FPs} can equally well be obtained with the simpler prescription in \cite{Eichhorn:2013isa}.

The elementary operators $\tr\left(\phi^{n_1}\,X^{m_1}\,...\right)...\tr\left(...\right)$ form a basis of theory space, i.e., a general action functional can be expanded as a linear combination of the elementary operators \footnote{The most general action functional is an arbitrary function of the matrix entries. However, since we use a vertex expansion to derive the RHS of the flow equation, we restrict ourselves to action functionals that can be expressed as linear combinations of the elementary operators. Notice also that this expansion is formal, i.e., we do not specify any norm in which this expansion is assumed to converge. The specification of such a norm is a very delicate problem, similar to summability of perturbation series.}. We thus have to find a projection prescription that allows us to extract the coefficients of the expansion of a general action functional in terms of elementary operators. This algorithm has to respect linearity of the expansion and it has to satisfy the elementary projection property: If the action functional is precisely one elementary operator then the expansion coefficient of this operator is $1$, while all other vanish.\footnote{This statement does of course require that elementary operators are linearly independent, which poses a restriction that we will discuss below.}

For practical calculations we will use the expansion $\left(\Gamma^{(2)}_N+R_N\right)^{-1}=\sum_{n=0}^\infty\,P^{-1}\left(-\,F[\phi]\,P^{-1}\right)^n$, where $P$ and $F[\phi]$ are fixed by $\Gamma_N^{(2)}[\phi]+R_N=:P+F[\phi]$ and $F[\phi\equiv 0]:=0$. This leads to a significant simplification, since this ensures that the RHS of the flow equation is ``polynomial'' in  $\phi$ whenever $\Gamma_N[\phi]$ is. For polynomial action functionals one can proceed as follows:

Each term on the RHS of the flow equation is of the form $T_n[\phi]=\frac {(-1)^n} 2 \textrm{tr}_{op}\left(P^{-1}\,\dot R\,\left(P^{-1}\,F[\phi]\right)^n\right)$, where we defined $\dot R = N \partial_N R_N$. 
We expand 
\be
T_n[\phi]:=\sum_{k=1}^\infty V_{n,k}^{a_1,b_1,...,a_{2k},b_{2k}}\phi_{a_1b_1}...\phi_{a_{2k}b_{2k}},
\ee
where 
\begin{equation}
 V_{n,k}^{a_1,b_1,...,a_{2k},b_{2k}}=\left.\frac{1}{(2\,k)!}\frac{\delta^{2k}\,T_n[\phi]}{\delta\phi_{a_1b_1}...\phi_{a_{2k}b_{2k}}}\right|_{\phi\equiv 0}.
\end{equation}
For instance, starting from an action of the form $\Gamma_N = {\rm tr} \phi^2+ \frac{g_4}4{} {\rm tr}\phi^4$ we will have $V_{n,k} \sim g_4^n \, \delta_{k,2n}$.
We will now use the fact that $\frac{\delta \phi_{ab}}{\delta \phi_{cd}}=\mathbb{I}$, where $\mathbb{I}$ is the appropriate  unit matrix, e.g., $\mathbb{I}= \frac{1}{2} (\delta_{ac} \delta_{bd}+\delta_{ad} \delta_{bc})$ \footnote{The symmetrization is model dependent, e.g., an unconstrained real model will have no symmetrization, a real symmetric model will require symmetrization in $a,b$ and complex matrix models require decomposition into real modes and subsequent (anti-) symmetrization if the model is Hermitian.}. Further, we choose an IR-suppression term that is an index dependent function times the appropriate symmetrization of $\delta_{..}\delta_{..}$. It thus follows that
\begin{equation}
 V^{a_1,b_1,...,a_{2k},b_{2k}}_{n,k}=\sum_i f_{n,k,i}^{a_1,b_1,...,a_{2k},b_{2k}}(a_1,...,b_{2k})\,\overbrace{\delta_{..,..}\,...\,\delta_{..,..}}^{\textrm{2k factors}},
\end{equation}
where each upper index of any $f_{n,k,i}^{a_1,b_1,...,a_{2k},b_{2k}}$ is contracted with an index of a Kronecker delta. We can thus write each $f_{n,k,i}^{a_1,b_1,...,a_{2k},b_{2k}}(a_1,...,b_{2k})$ as a function of the indices $g(a_1,...,b_{2k})$ times a contraction pattern (which is given by the Kronecker-deltas). We now perform a Taylor-expansion of the $g$ around vanishing index
\begin{equation} 
 \begin{array}{rcl}
   g(i_1,...,i_{4k})&=&\left.g\right|_{\vec i\equiv 0}+\left.\frac{\partial \, g}{\partial\, i_j}\right|_{\vec i\equiv 0}\,i_j\\
    &&+\frac 1 2 \left.\frac{\partial^2\,g}{\partial\, i_j\,\partial\, i_k}\right|_{\vec i\equiv 0} \, i_j\,i_k + ...\,\,\\
   &=&g^0+g^1_{i_1}\,i_1+\frac 1 2 \, g^2_{i_1,i_2}\,i_1\,i_2+...
 \end{array}
\end{equation}
The expansion coefficients $g^k_{i_1,...,i_k}$ combined with the contraction pattern $\delta_{i_.,i_.}...\delta_{i_.,i_.}$ can be identified as being generated by a unique product of traces of matrix products of $\phi$- and $X$-matrices: The contraction patters determine how many $\phi$'s appear in each trace and the dependence of $g^k_{i_1,...,i_k}$ on the indices $i_1,...$ tell us at which positions what power of the $X$-matrices have to be inserted. To project onto the symmetric operators with no $X$-insertions, we would only consider the $g^0$ term.

This procedure allows us to uniquely expand the RHS of the flow equation in terms of the elementary monomials $\textrm{Tr}\left(\phi^{n_1}\,X^{n_2}...\right)...\textrm{Tr}\left(\phi^{m_1}\,X^{m_2}...\right)$ that we use as the coordinate basis for our theory space. Two subtleties associated with the use of this basis should be noted:
\begin{enumerate}
 \item Commutativity of the product of traces and cyclicity of the trace imply the monomials $\textrm{Tr}\left(\phi^{n_1}\,X^{n_2}...\right)...\textrm{Tr}\left(\phi^{m_1}\,X^{m_2}...\right)$ are not simply labeled by the arrays of integers $((n_1,n_2,...),...,(m_1,m_2,...))$, but by equivalence under all permutations of blocks and cyclic permutations by an even number of steps of numbers within a block. This means that we have to label the coordinate basis of the theory space, i.e., the coupling constants, by fixing a unique representative in each equivalence class.
 \item The regulator is not an analytic function of the indices due to the Heaviside function. This Heaviside function has however no observable effect when the effective action is probed with ``IR''-degrees of freedom (i.e., matrices $\phi$ that have only the upper left $N\times N$ components nonvanishing). Thus, for effective IR field theory, one can use the above identification of field monomials, since the Taylor expansion of the functions $g$ around vanishing index is ``blind'' to the Heaviside function.
\end{enumerate}

\section{Renormalization Group flow and gauge symmetry}\label{sec:RGSym}
The above procedure would allow us to derive the RG flow in the extended, non-symmetric theory space. Since our model is symmetric under $U(N)$, and the symmetry-breaking is only introduced by the regulator, there is a Ward-identity that will impose a nontrivial constraint on the RG flow in the extended theory space.
The action of a Hermitian pure matrix model is invariant under unitary transformations which act on the field in the form
\begin{equation}
 \phi \mapsto O^T\,\phi\,O = \phi+\epsilon\,[\phi,A]+\mathcal O(\epsilon^2),
\end{equation}
where $A$ is the generator of an infinitesimal symmetry transformation. The functional measure and the bare action of a pure matrix model are invariant under unitary transformations, but the change of the regulator term is
\begin{equation}
 \mathcal G_\epsilon\,\Delta_N\,S = \epsilon\,\textrm{Tr}\left(\phi\,[A,R_N]\,\phi\right),
\end{equation}
where we denoted the change of a functional $F$ under an infinitesimal gauge transformation by $\mathcal G_\epsilon\,F$. Thus, the effective average action satisfies the scale dependent Ward-Takahashi identity (WTI)
\begin{equation}\label{equ:WTI}
 \mathcal W_N\,\Gamma_N=\mathcal G_\epsilon\,\Gamma_N-\textrm{tr}_{\small{op}}\left(\frac{[A,R_N]}{\Gamma_N^{(2)}+R_N}\right)=0.
\end{equation}
It follows form the standard argument see, e.g., \cite{Gies:2006wv}, that the RG-evolution of an initial condition that satisfies the initial WTI $\mathcal W_N\,\Gamma_N=0$ at an initial scale $N$, will satisfy the evolved WTI $\mathcal W_{N^\prime}\,\Gamma_{N^\prime}=0$ at a scale $N^\prime$. Hence, if we want to implement gauge symmetry, i.e., if we require the usual WTI $\Gamma_\epsilon\,\Gamma=0$ to hold, then we have to impose that satisfies $\Gamma_N$ the scale dependent WIT to ensure that $\Gamma=\lim_{N\to 0}\Gamma_N$ satisfies the usual WTI.

It is important to notice that the scale dependent WTI can not be solved by a $\Gamma_N$ that respects the usual gauge symmetry $\mathcal G_\epsilon\,\Gamma_N=0$, because the second term of (\ref{equ:WTI}) does not vanish unless $N=0$. Hence, to implement gauge symmetry in the flow, one is forced to ``contaminate'' $\Gamma_N$ by turning on just the right amount of couplings for symmetry-breaking operators. Conversely, to implement gauge symmetry, we have to restrict the search for RG fixed points to solutions of the scale dependent WTI. 
\subsection{Tadpole Approximation}\label{sec:tadpole}
Restricting the search for RG-fixed points to solutions of the scale dependent WTI will be dealt with in future work. For the present paper, we make the following important observation: If we assume that the sought-for fixed point lies at small values of the couplings, then we can approximate the $\beta$ functions by the first order in the vertex expansion: Assuming that combinatorial factors in the loop diagrams are $\mathcal{O}(1)$, and the fixed-point value of all couplings is $\sim \epsilon<1$, then the $n$-vertex diagram is suppressed by a factor $\epsilon^{n-1}$ in comparison to the tadpole diagram. Accordingly it is a self-consistent approximation to take into account only tadpole diagrams, if the corresponding fixed-point values indeed turn out to satisfy our requirement. We now make the following central observation: As all vertices arising within a truncation consisting of $U(N)$ invariants are themselves $U(N)$ invariant, symmetry-breaking operators cannot be generated by the tadpole diagram. Any nontrivial index-dependence on the RHS of the flow equation can always be shifted away, as there is no nontrivial index dependence in the vertex. This is completely analogous to the case of, e.g., standard $\lambda \phi^4$ theory: As the vertex proportional to $\lambda$ is momentum-independent, the tadpole diagram cannot generate a momentum-dependent operator. Thus no non-trivial wave-function renormalization is generated from the tadpole diagram $\sim \lambda$. In our case, a non-trivial index-dependence is analogous to a non-trivial momentum dependence. 

This reasoning can be applied to the vertex expansion of the RHS of the flow equation as well as the vertex expansion of the second term of the scale dependent WTI (\ref{equ:WTI}). We conclude that the scale-dependent WTI is solved by a $U(N)$-symmetric $\Gamma_N$ in the tadpole approximation and, conversely, that the tadpole approximation to the RG flow preserves $U(N)$-symmetry. 

One might now wonder whether the functional Renormalization Group will be of use to uncover the double-scaling limit in higher-dimensional tensor models. If the critical value of the coupling would lie at a large value, the tadpole approximation would not be applicable. Here it is crucial that the critical value of the coupling corresponds to the radius of convergence of the perturbative expansion, and as such is guaranteed to lie at values much smaller than one. Accordingly, the use of the tadpole approximation is justified to explore the double-scaling limit in higher-dimensional tensor models. An added benefit lies in the fact that the evaluation of the $\beta$ functions in a large truncation is simplified considerably, if we restrict ourselves to the tadpole approximation. Thus we are confident that our method will also allow us to successfully tackle higher-dimensional tensor models.
\section{$\beta$-functions and fixed points}\label{sec:FPs}
The considerations of the previous section suggest that the tadpole approximation to the $\beta$-functions will improve the results of \cite{Eichhorn:2013isa} for fixed points with small values of the couplings. We now confirm this suggestion and in this course also uncover the multicritical fixed points with the FRG for the first time.
\subsection{Single-trace approximation}
As a first step, let us reconsider the single-trace truncation studied in \cite{Eichhorn:2013isa}, which is
\begin{equation}
 \Gamma_k = Z_{\phi} \tr \phi^2 + \sum_{n=2}^{{n}_\textrm{max}} \frac{g_{2n}}{2n} \tr \left(\phi^{2n}\right),
\end{equation}
where we take $n_\textrm{max}=7$ in accordance with \cite{Eichhorn:2013isa}.
Employing eq. 63 in \cite{Eichhorn:2013isa}, restricting ourselves to tadpole diagrams,  we then obtain a set of beta functions as follows:
\begin{eqnarray}
  \eta &=& 2 \,g_4 \,x,\\
 \beta_{{2n}}&=&\left((n-1) + n \,\eta \right)g_{2n} - 2\,n\, x\, g_{2(n+1)},
\end{eqnarray}
where the first term in the beta functions arises from the canonical dimensionality of the couplings and $\eta = - N \partial_N \ln Z_{\phi}$. We have set $[\dot{R}P^{-2}] =x$ in order to study the scheme dependence of our results.

Here, we also neglect the term $\sim \eta$, that is generated by $\partial_N R_N$ on the right-hand-side of the flow equation.  

We then obtain a set of fixed points and critical exponents listed in tab.~\ref{singletraceFPs}.

\begin{widetext}

\begin{table}[!here]
\begin{tabular}{cccccc|cccccc}
 $g_4$ & $g_6$ & $g_8$ & $g_{10}$ & $g_{12}$ & $g_{14}$ & $\theta_1$ &$\theta_2$ &$\theta_3$ &$\theta_4$ & $\theta_5$ & $\theta_6$ \\ \hline\hline
0 & 0 & 0 & 0 & 0 & 0& -1 & -2 & -3 & -4& -5 & -6\\ \hline
$\frac{-1}{4 x}$ & 0 & 0 & 0 & 0 & 0 & 1 &$-\frac{1}{2}$ & -1 & $-\frac{3}{2}$ & -2 & -$\frac{5}{2}$ \\ \vspace{-0.3cm}\\ \hline
$\frac{-1}{3x}$ & $\frac{1}{36 x^2}$ & 0 &0 &0 &0 & 1 & $\frac{2}{3}$ & $- \frac{1}{3}$ & $- \frac{2}{3}$ & -1 & $- \frac{4}{3}$\\ \vspace{-0.3cm}\\ \hline
$\frac{-3}{8 x}$ & $\frac{3}{64 x^2}$ & $- \frac{1}{512 x^3}$ & 0 & 0 & 0 & 1 &$\frac{3}{4}$ & $\frac{1}{2}$ & $-\frac{1}{4}$ & $- \frac{1}{2}$ & $- \frac{3}{4}$ \\  \vspace{-0.3cm}\\\hline
$\frac{-2}{5x} $& $\frac{3}{50 x^2}$ & $\frac{-1}{250 x^3}$ & $\frac{1}{10^4 x^4}$ & 0 & 0 & 1 & $\frac{4}{5}$ & $\frac{3}{5}$ & $\frac{2}{5}$ & $-\frac{1}{5}$ & $- \frac{2}{5}$  \\  \vspace{-0.3cm}\\\hline
$\frac{-5}{12 x}$ & $\frac{5}{72 x^2}$ & $- \frac{5}{864 x^3}$ & $\frac{5}{20736 x^4}$& $-\frac{1}{248832 x^5}$ & 0 & 1 & $\frac{5}{6}$ & $\frac{2}{3}$ & $\frac{1}{2}$ & $\frac{1}{3}$ & $- \frac{1}{6}$  \\  \vspace{-0.3cm}\\\hline
$\frac{-3}{7 x}$ & $\frac{15}{196 x^2}$ & $-\frac{5}{686 x^3}$ & $\frac{15}{38416 x^4}$ & $-\frac{3}{268912 x^5}$ &$\frac{1}{7529537 x^6}$ & 1 & $\frac{6}{7}$ & $\frac{5}{7}$ & $\frac{4}{7}$ & $\frac{3}{7}$ & $\frac{2}{7}$ 
\end{tabular}
\caption{\label{singletraceFPs} We show fixed points and critical exponents that we obtain in a single-trace truncation including all couplings up to $g_{14}$. We include only tadpole diagrams, and parameterize $\dot{R}P^2 =x$.}
\end{table}

\end{widetext}

The first fixed point is the Gau\ss{}ian fixed point, where the critical exponents equal the canonical scaling dimensionality of the couplings.

The second fixed point corresponds to the well-known double-scaling limit, with one relevant direction. As in \cite{Eichhorn:2013isa}, this first approximation yields a critical exponent $\theta=1$, instead of the analytically known exact value $\theta = \frac{4}{5}$.

\subsubsection*{Multicritical Points}
All other fixed points correspond to multicritical points of increasing order $m$. They show the well-known pattern of alternating signs for the couplings, corresponding to stable/unstable potentials. As expected for the $m$th multicritical point, $m-1$ relevant directions exist: The largest critical exponent corresponds to the pure-gravity case \cite{Brezin:1992cy}. The next critical exponents are expected to be $\theta_{m-1} = \frac{2}{m+\frac{1}{2}}$. We obtain, similarly to \cite{Brezin:1992yc}, $\theta_{m-1} = \frac{2}{m}$.

The confirmation of the existence of the multicritical points within the FRG approach to matrix models is one of the main new findings of this paper.

Moreover, one can find analytic expressions for the multicritical points in the tadpole approximation to the single-trace truncation as follows: assume that the $g_{2n}$  vanish for all $n>m$, so the $\beta_{2n}$ vanish for all $n>m$, while $\beta_{2m}=0$ and $\eta=2\,x\,g_4$ imply
\begin{equation}\label{equ:etag4}
 \eta=\frac{1-m}{m},\quad\,g_4=\frac{1-m}{2\,m\,x}.
\end{equation}
The vanishing of the remaining $\beta_{2n}$ give the linear recursion relation $g_{2n+2}=\left(\frac{n-1}{n}-\frac{m-1}{m}\right)\frac{g_{2n}}{2\,x}$. For the initial condition (\ref{equ:etag4}) the solution is, in agreement with tab.~\ref{singletraceFPs}, given by
\begin{equation}\label{equ:STrecursionSOL}
  g_{2n}=\frac{(1-m)_{n-1}}{(n-1)!}(2\,m\,x)^{1-n}, 
\end{equation}
where we used the Pochhammer symbol $(a)_n=a(a+1)...(a+n)$.

After inserting the $k$-th multicritical value $\eta=\frac{1-k}{k}$ into the tadpole approximation to the vertex expansion, one finds that the Jacobian 
\begin{equation}
  \frac{\partial\,\beta_{2n}}{\partial g_{2m}}=\delta_{n,m}\left((n-1)-\frac{n}{m}(m-1)\right)-2\,\delta_{n+1,m} \,n\,x 
\end{equation}
is triangular and independent of the couplings $g_{2n}$. We can thus recover the positive critical exponents from the diagonal entries of the Jacobian
\begin{equation}
 \theta^{(m)}_n=\frac n m,\quad\textrm{ where: }n=2,...,m
\end{equation}
in agreement with tab.~\ref{singletraceFPs}. (Notice that, for simplicity of presentation, we treated $\eta$ as a constant and not as a function of $g_4$, when we calculated the critical exponents, which turns out not to have an effect on the result in the single trace approximation.)
\subsubsection*{Universality}
We observe that, although the $g_{n\, \ast}$ depend on $x$, the critical exponents do not. Thus these values are universal, i.e., independent of the choice of regularization scheme. The dimensionful couplings themselves are not universal, i.e. $g_{n\, \ast} = g_{n\, \ast}(x)$. However, we can also form universal (dimensionless) ratios of couplings, such as $g_4^2/g_6$, which accordingly are independent of $x$. For instance, the second multicritical point, where $g_8=0$, has $g_4^2/g_6 = 4$, which is in reasonable agreement with the exact $g_4^2/g_6 = \frac{10}{3}$, and in fact corresponds exactly to the value in \cite{Brezin:1992yc}.

Note that to obtain the value $g_{\ast} = g_c = -\frac{1}{12}$ for $m=2$, we would have to set $x=3$, which clearly shows that fixed-point values are non-universal. This is expected as in our setting they carry a non-trivial scaling dimensionality with $N$. In the same way that fixed-point values for dimensionful couplings cannot be universal in standard quantum field theories, no such universality is expected in our case.

A subtle difference to standard quantum field theories arises, as the notion of scale that we introduce here disappears, once the integration over all quantum fluctuations has been completed: In the limit $N \rightarrow \infty$, there is no other quantity left in the matrix models that would still set a scale. This is a major difference to standard quantum field theories: There, two distinct notions of scale exist: One is a Renormalization Group scale $\mu$, that decides which quantum fluctuations have been integrated out. The other is a model-inherent momentum scale, which enters the operators of the model, such as, e.g., the kinetic term. Since both scales are momentum scales, the dimensionality assigned to couplings using the model-inherent momentum, or the ''external" RG scale, agrees. In particular, a nontrivial notion of dimensionality remains in the limit $\mu \rightarrow \infty$. This is different in the case of the matrix model, where non-trivial scaling dimensions do not exist once $N \rightarrow \infty$. The couplings become dimensionless in that limit. This observation should explain, why our fixed-point values are nonuniversal, whereas other methods that are used to derive the double-scaling limit in matrix models yield a universal number for $g_{4\, c}$.
\subsection{Multi-trace truncation}
We now investigate a truncation that takes into account all operators up to a fixed dimensionality. In particular, we include $g_4, \dots, g_{12}$, $g_{22}, \dots,g_{28}$, $g_{44}, g_{46}$ and $g_{222}, g_{224}$. Again, we restrict ourselves to tadpole diagrams.  According to our reasoning in sec.~\ref{FRG}, the canonical dimensionality provides a useful organizing principle for the couplings in theory space. We thus expect that this multitrace truncation should show improved results over the single-trace truncation, as it will consistently take into account \emph{all} operators up to a given dimensionality.

The interesting double- and triple trace operators are contained in the following truncation
\begin{widetext}
\begin{equation}
 \begin{array}{rcl}
 \Gamma_N&=&\frac{Z_{\phi}}{2} \tr\left(\phi^2\right) + \sum_{n\ge 2}\frac{g_{2n}}{2}\tr\left(\phi^{2n}\right)+\sum_{n\ge 1}\frac{g_{2,2n}}{2}\tr\left(\phi^2\right)\,\tr\left(\phi^{2n}\right) \\
          &&+ \sum_{n\ge 2}\frac{g_{4,2n}}{2}\tr\left(\phi^4\right)\,\tr\left(\phi^{2n}\right)
          +\sum_{n\ge 1}\frac{g_{2,2,2n}}{2}\tr\left(\phi^2\right)^2\,\tr\left(\phi^{2n}\right).
  \end{array}
\end{equation}
The tadpole approximation to the beta functions in this truncation is
\begin{eqnarray}
  \eta&=&2\,x\,\left(g_4+g_{2,2}\right)\\
  \beta_{2n}&=&\left((n-1)+n\,\eta\right)g_{2n}-n\,x\left(2\,g_{2n+2}+g_{2,2n}\right)\\
  \beta_{2,2}&=&2\left(1+\eta\right)g_{2,2}-x\left(g_6+4\,g_{2,4}+3\,g_{2,2,2}\right)\\
  \beta_{2,4}&=&3\left(1+\eta\right)g_{2,4}-2\,x\left(g_8+3\,g_{2,6}+4\,g_{4,4}+g_{2,2,4}\right)\\
  \beta_{2,2n}&=&(n+1)\left(1+\eta\right)g_{2,2n}-2\,x\left(g_{2n+4}+(n+1)g_{2,2n+2}+2\,g_{4,2n}+g_{2,2,2n}\right),\,\, n\geq3 \\
  \beta_{4,4}&=&4\left(1+\eta\right)g_{4,4}-x\left(g_{10}+6\,g_{4,6}\right)\\
  \beta_{4,2n}&=&(n+2)\left(1+\eta\right)g_{4,2n}-x\left(2\,g_{2n+6}+(2n+2)g_{4,2n+2}\right), \, \, n\geq3\\
  \beta_{2,2,2}&=&\left(4+3\,\eta\right)g_{2,2,2}-x\left(3\,g_{2,6}+4\,g_{2,2,4}\right)\\
  \beta_{2,2,4}&=&\left(5+4\,\eta\right)g_{2,2,4}-x\left(g_{2,8}+3g_{4,6}+6\,g_{2,2,6}\right)\\
  \beta_{2,2,2n}&=&\left((n+3)+(n+2)\eta\right)g_{2,2,2n}-x\left((2\,n+4)g_{2,2n+4}+(2n+2)g_{2,2,2n+2}\right),\\
\end{eqnarray}
\end{widetext}
where the exceptions arise due to additional symmetry factors (e.g., the derivative of $\tr \phi^4 \tr \phi^4$ has an additional factor of 2 when coupling into $g_{2,4}$, than the derivative of $\tr \phi^4 \tr \phi^6$ when coupling into $g_{2,6}$. )

\begin{figure}[!here]
\includegraphics[width=\linewidth]{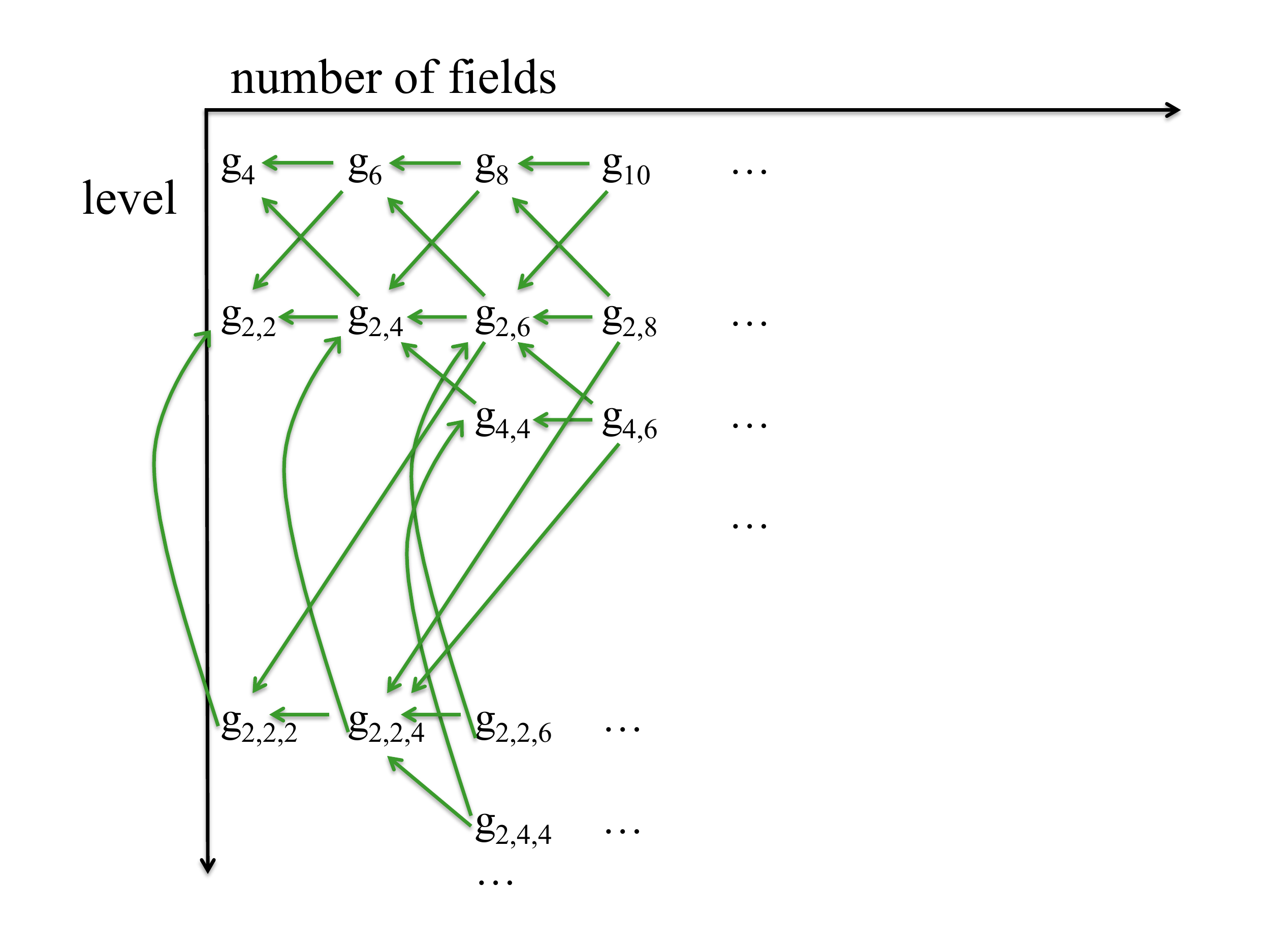}
\caption{\label{levelplot} We schematically show the structure of the flow equation. We indicate tapole diagrams by arrows. Clearly only neighbouring levels couple into each other's beta functions. Note that at higher order in the traces, the levels get more complicated, e.g., there exist two-trace terms of the form $g_{2,i}, g_{4,i}, g_{6,i}...$.}
\end{figure}

The general structure has already been discussed in \cite{Eichhorn:2013isa}, and is as follows, cf.~fig.~\ref{levelplot}: At leading order in $1/N$, only a tadpole diagram of $g_{2,i_1, \dots, i_n}$ can couple into $g_{i_1,\dots , i_n}$. Our truncation to tadpole diagrams is therefore not a truncation, but already the full result, when it comes to the back-coupling of higher orders in the number of traces into the lower-order beta functions.

Furthermore, an $n$-trace operator can only generate an $(n+1)$-trace operator through a tadpole diagram, as the contraction on the right-hand side of the flow equation does not generate more than one additional traces: At most, the structure of $\Gamma^{(2)}$ and the subsequent contraction permits to split one of the $\tr \phi^i$-terms into two new, $\tr \phi^{i-j} \tr \phi^j$ terms. 

We thus have a structure where only neighbouring ``levels" (i.e., numbers of traces) are coupled in the beta functions. Thus, already the three-trace terms affect the single-trace terms only indirectly.
Additionally, many contributions that could be possible if one only counts the number of fields, are disallowed because of the trace-structure of the flow equation: For instance, just counting the number of fields, one could expect a coupling of $g_{4,4}$ into $\beta_{g_{2,2,2}}$. This contribution does not exist, as the second derivative of $\tr \phi^4 \tr \phi^4$ cannot generate a term that separates into 3 traces with two fields each upon contraction with the propagator.

Beyond the tadpole level, two-vertex diagrams can ``span" a larger number of levels, as combinations of vertices with $i$ traces and vertices with $i+2$ traces can also couple into operators with $i+2$ traces.

For our explicit solution of the fixed-point equations, we only include up to $g_{12}$. For consistency, this implies that we have to take into account up to three-trace operators.
We thus expect to find the first 5 multicritical points. It turns out, that a subset of the multitrace couplings is also nonvanishing at these multicritical points, and that they have an important effect on the critical exponents, cf. tab.~\ref{multitracetab}.

\begin{widetext}

\begin{table}[!here]
\begin{tabular}{ccccc|cccc|cc|cc|ccccc}
$g_4$ & $g_6$ & $g_8$ & $g_{10}$ & $g_{12}$ & $g_{22}$ & $g_{24}$ &$g_{26}$& $g_{28}$ & $g_{44}$ & $g_{46}$ & $g_{222}$ & $g_{224}$ & $\theta_1$& $\theta_2$  & $\theta_3$ & $\theta_4$ & $\theta_5$ \\ \hline
 $-\frac{1}{4x}$ & 0 & 0 &0 &  0 & 0& 0 & 0& 0 & 0& 0 & 0&0& 1 & -1/2 & -1 & -1 & -3/2\\\vspace{-0.3cm}\\\hline
$-\frac{8}{21x}$ & $\frac{2}{63 x^2}$ & 0 & 0 & 0 & $\frac{1}{21 x}$ & 0 &0 &0 & 0 &0 & 0&0& 0.82 & 0.82 & -0.33 & -0.64 & -0.67 \\\vspace{-0.3cm}\\\hline
$- \frac{51}{116 x}$ & $\frac{27}{464 x^2}$ & $- \frac{9}{3712 x^3}$ & 0 & 0 & $\frac{15}{232 x}$ & $-\frac{3}{464 x^2}$ & 0 & 0 & 0 & 0 & 0 & 0 & 0.80 & 0.80 & 0.61 & -0.25 & -0.5   \\\vspace{-0.3cm}\\\hline
$\frac{-16448}{34535 x}$ & $\frac{13392}{172 675 x^2}$ & $ \frac{-4608}{863375 x^3}$ & $\frac{576}{4316875 x^4}$ & 0 & $\frac{2634}{34535 x}$ & $ \frac{-2112}{172675 x^2}$ & $\frac{288}{863375 x^3}$ & 0 & $\frac{108}{172675 x^2}$ & 0 & $\frac{144}{863375 x^3}$ & 0 & 0.81 & 0.81 & 0.68 & 0.45 & -0.2\\\vspace{-0.3cm}\\\hline
\end{tabular}
\caption{\label{multitracetab}We show the first four fixed points and critical exponents that we obtain in a multi-trace truncation. We include only tadpole diagrams, and parameterize $\dot{R}P^{-2} =x$. A number of further fixed points that we obtain is not shown, as they do not correspond to any known analytical solution and are thus most probably artifacts of the truncation.}
\end{table}

\end{widetext}

Most importantly, the largest critical exponent which corresponds to the pure-gravity critical exponent turns out to be $\theta \sim 0.8$ for all but the first fixed point. We thus observe that the inclusion of multi-trace terms is a crucial step towards quantitative precision in matrix models. While the sub-leading critical exponents at the multicritical points deviate more significantly from the exact values, the leading critical exponent only deviates by a few percent from the exact result $\theta=0.8$. This is a major step forward from the investigations in \cite{Eichhorn:2013isa}. Already for the pure-gravity fixed point, our result constitutes an improvement over our previous result in a multi-trace truncation, as well as the considerable improvement over the perturbative calculation in \cite{Ayala:1993fj}. 

For the fixed-point values, we observe that universal combinations such as $\frac{g_4^2}{g_6}$ depart further from the exact result than in the single-trace approximation. We attribute this to the fact that multi-trace operators at the same order of the fields are nonvanishing, e.g., $g_{2,2} \neq0$ at the $m=3$ multicritical point.
\subsubsection*{Analytic Solution}
If the initial condition to the flow is such that all couplings that correspond to operators with more than $2k$ fields vanish, then the tadpole approximation to the RG flow will preserve this condition, because a tadpole diagram from a truncation with $2k$ fields will generate operators with at most $2k-2$ fields. It is thus, in the tadpole approximation, consistent to search for fixed points at which the couplings for all operators with more than $2k$ fields vanish. For such a fixed point search, one can employ the following strategy:
\begin{enumerate}
 \item There will be a finite number (the number of distinct integer partitions of $k$) of beta functions for the operators with $2k$ fields, which are of the form $\beta_{a}=(\textrm{dim}(a)+k\,\eta)g_{a}$. These imply that either that $\eta=-\frac{\textrm{dim}(a)}{k}$ for one $a$ and all other couplings with $2k$ fields vanish just as in tab.~\ref{multitracetab},
 or that all $g_{...}$ vanish. The second case corresponds to a fixed point with at most $2k-2$ field and is thus the same case, but with different $k$. 
 \item The remaining beta functions are of the form $\beta_{b}=(\textrm{dim}(b)-\textrm{dim}(a)\frac{2n}{k})g^{(2n)}_b+x(\textrm{linear in }g^{(2n+2)})$, where the superscript bracket denotes the number of fields in the corresponding operator. Vanishing of the beta functions thus gives the recursion relation $g^{(2n)}_b=\frac{\textrm{dim}(a)\frac{2n}{k}-\textrm{dim}(b)}{x(\textrm{linear in}g^{(2n+2)}}$.
 \item One now sets $g_a=\alpha$ and uses the recursion relations to derive all coupling constants with $2k-2$ fields, $2k-4$ fields, ... and $4$ fields as functions of $\alpha$. This provides in particular $g_4(\alpha)$ and $g_{2,2}(\alpha)$. Notice that the recursion relation is linear, which implies that $g_b(\alpha)=\alpha\,g_b(1)$.
 \item The anomalous dimension then implies that $2\,\alpha\,x(g_4(1)+g_{2,2}(1))=-\frac{\textrm{dim}(a)}{k}$, which shows that $\alpha=-\frac{\textrm{dim}(a)}{2\,k\,x\,(g_4(1)+g_{2,2}(1))}$.
\end{enumerate}
A technical difference between the single- and multitrace truncation is that it is easy to find the solution of the singletrace recursion relation as a function of $k$ (see \Eqref{equ:STrecursionSOL} above), while we where unable to express the solution of the multitrace recursion relation as a function of $k$.
\subsection{Two-vertex contributions}
The next logical step is to include two-vertex contributions to the beta functions, which provide the leading order corrections to the tadpole approximation for fixed points with small couplings. These contributions do, as we explained above, imply that the solution to the Ward identity requires the inclusion of non-symmetric operators. In the explicit symmetric truncations that we considered so far, it turns out that the inclusion of two-vertex contributions has far-reaching effects: We observe, that none of the multicritical points can be found after the inclusion of these additional terms, and only the pure-gravity double-scaling limit remains. One of course expects that the inclusion of two-vertex diagrams will lead to an improvement once one restricts the flow to a consistent approximation to the solution of the Ward identity. The test of this expectation goes beyond the scope of this paper and will be investigated in a future paper.

\section{Connection to continuum $\beta$ functions for gravity and the asymptotic safety scenario}\label{connection}

In research on quantum gravity, many approaches exist in parallel and it is a priori unclear whether they are in any way related. In particular, the continuum approach based on a quantum field theory for the metric, known as the asymptotic safety scenario \cite{Weinberg:1976xy, Weinberg:1980gg, Reuter:1996cp}, for reviews see, e.g., \cite{AS_reviews}, and approaches based on a discretization of geometry, such as matrix or tensor models, differ in many aspects. The fundamental variables are taken to be different (metric versus matrices/tensors), the symmetries differ (diffeomorphism symmetry versus $U(N)$ symmetry), and the Renormalization Group flow is formulated with respect to two completely different notions of scale (defined with respect to a fiducial background metric versus matrix/tensor size). Nevertheless, one could expect the following scenario, see, e.g., \cite{Oriti:2013jga,Rivasseau:2014ima} and also a related discussion in \cite{Percacci:2010af}, where the interacting fixed point underlying asymptotic safety is related to a 
phase transition from the ``pregeometric" to the geometric phase of a tensor model/group field theory.

To be more specific,  the tentative non-trivial continuum limit in matrix/tensor models, signaled by an interacting fixed point, is characterized by a set of critical exponents. In a more physical sense, that limit can also be interpreted as a phase transition from a pre-geometric phase, where no notion of the metric exists, to a geometric phase with a non-vanishing expectation value of the metric. The approach to this continuous (i.e., second order or higher) phase transition is described by the critical exponents see, e.g., \cite{Ambjorn:2011cg} for a recent example in Causal Dynamical Triangulations.

In contrast, the continuum approach known as asymptotic safety is based on the existence of a non-vanishing metric, and cannot easily describe the phase with vanishing expectation value of the metric, see, however, \cite{Reuter:2008qx}. At very high momentum scales, the scaling of operators is determined by critical exponents of an interacting fixed point of the Renormalization Group flow. 

These two scenarios can be interpreted as two sides of the same picture, where the same phase transition -- from a pre-geometric phase to a geometric phase -- is approached from the two sides: Matrix/tensor models are extremely well-adapted to describe the pregeometric phase, and the approach to that phase transition. On the other hand, the physics of the geometric phase is then more straightforwardly accessed in the continuum quantum field theory setting. 
For this connection between the two settings to hold, the critical exponents calculated on both sides should agree. Further evidence for such a connection between continuum and discrete quantum gravity could be provided by observables, such as, e.g., the spectral dimension, which has been found to equal two in the UV in both discrete \cite{Ambjorn:2005db} as well as continuum settings \cite{Lauscher:2005qz, Horava:2009if}.

In the following, we will review explicitly, how the fixed point in matrix models and the corresponding critical exponent is related to the beta function of $2+\epsilon$ dimensional continuum quantum gravity \cite{Weinberg:1980gg, twoplusepsilon}. This provides a simple example of the relationship between the continuum setting (i.e. the asymptotic safety scenario) and the matrix/tensor model setting. Whether a similar relationship exists in higher dimensions, remains to be investigated.

The well-known geometric picture that underlies the matrix model approach to quantum gravity is the discrete approximation of the geometry of a compact\footnote{Strictly speaking we demand compact without boundary.} Riemann surface by tesselations with indistinguishable building blocks, e.g., be equilateral triangles or squares. A continuum geometry is attained in the limit in which the number $N$ of elementary building blocks diverges while their individual size is rescaled by scaling the lattice constant $a$ as $a/N^{\frac 1 2}$, so the total area $A$ of the Riemann surface remains unchanged.

The discrete approximation at finite $N$ defines a regularized measure for the functional integral of two-dimensional Euclidean quantum gravity as the sum over all tesselations. For this one observes that the Einstein-Hilbert action in two dimensions consists of a cosmological constant term proportional to the total area $A$ and a topological term proportional to the Euler characteristic $\chi$. Both quantities possess simple expressions in terms of the tesselation: the total area is $N\,a^2$ and the Euler characteristic is number of vertices $V$ - edges $E$ + faces $F$. The aim of this approach is to define the functional measure by taking the continuum limit $a\to 0$, $a/N^{\frac 1 2}=const.$.

The discrete partition function at finite $N$ can be represented as a matrix model. The underlying observation is that the Feynman graphs of a matrix model possess an interpretation in terms of tesselations, which arise as the dual to the ribbon graphs of the matrix model. Starting from the matrix- model action $S=N \tr (-\frac{1}{2} \phi^2 + g \phi^4)$, one sees that each closed loop contributes a factor $N$ due to the summation of a free index, while the Feynman rules assign each vertex a factor $N$ and each propagator a factor $N^{-1}$. Thus, we obtain a factor $N^{V-E+F}=N^\chi$ for each Feynman diagram.

It follows that in matrix models, the matrix size $N$ is related to the bare Newton coupling by 
\be
N = e^{\frac{1}{4\, G_0}},\label{NewtonG}
\ee
as $\frac{1}{4 G_0}$ is the prefactor of $\chi$ in the action.

Furthermore, the relation 
\be
g_4 = e^{-\Lambda a^2}
\ee
holds for the dynamical triangulation, see, e.g., \cite{Ambjorn:1994yv}. This follows from the fact that each configuration is weighted by $e^{- \Lambda A} = e^{-\Lambda a^2 n}$, where $n$ is the number of squares. On the other hand, each discrete configuration, corresponding to a Feynman diagram, is weighted by $g_4^n$. We then translate to a dimensionless cosmological constant, $\lambda = a^{-2} \Lambda$.
 
Accordingly, the double-scaling-limit for pure gravity translates into the requirement \footnote{The following results were communicated to us by Jan Ambjorn.}:
\be
\left( g_4-g_{4\, c}\right)^{5/4} N = \left( e^{- \lambda} - e^{- \lambda_c}\right)^{5/4} e^{\frac{1}{4 G_0}} = \rm const.
\ee
This establishes the correspondence between the double scaling limit and the continuum limit.  We see that taking the lattice spacing $a \rightarrow 0$, while physical quantities, such as the renormalized cosmological constant, $\Lambda_R$, are held fixed, i.e., 
$\Lambda_R  = a^{-2} (\lambda - \lambda_c)$ is equivalent to taking the double scaling limit of the matrix model.

For $\lambda- \lambda_c \ll 1$, we then obtain that
\be
\left(\Lambda_R a^2\right)^{5/4} e^{\frac{1}{4  G_0}} = \rm const.\label{GLambda}
\ee
Taking the lattice spacing to zero then requires us to adjust the bare Newton coupling $G_0$ appropriately, i.e., $G_0 = G_0(a)$.  
It is then straightforward to derive the scale-dependence of $G_0$ from \Eqref{GLambda}:
\be
- a \partial_a G_0(a) = -10 \,G_0^2.
\ee
Since $a \partial_a = - \mu \partial_{\mu}$ where $\mu$ is a Renormalization Group momentum scale, we finally obtain
\be
\beta_G = \mu \partial_{\mu} G= - 10 \,G^2.
\ee
Note that the numerical value of the coefficient depends of course on whether the action is defined as $\frac{R}{G}$, or $\frac{R}{16 \pi G}$.
Let us comment on the connection with continuum field theory: In $2+\epsilon$ dimensions, where quantum gravity based on the Einstein-Hilbert action on a fixed topology is no longer trivial, one can obtain a nontrivial beta function form a standard quantum field theory calculation \cite{twoplusepsilon}. Its main feature is a term $\sim G^2$ with a negative sign. Together with the term arising from a nontrivial dimensionality in $d= 2 + \epsilon$, this term is responsible for the existence of a UV attractive interacting fixed point. 

Here, we have a similar result, with a term $\sim G^2$ with a negative sign, which corresponds to asymptotic freedom. As a difference to the results in \cite{Weinberg:1980gg, twoplusepsilon}, it arises in $d=2$. Crucially, it does not follow from a simple scaling limit, where $g_4 \rightarrow g_{4\, c}$. Instead, it originates from the double-scaling limit, where all topologies with higher Euler character contribute. One could thus interpret the existence of asymptotic freedom in $d=2$ dimensional quantum gravity as stemming from topological fluctuations. Most interestingly, going to $d=2+ \epsilon$, a term $\epsilon \,G$ will arise from the canonical dimensionality of the coupling. The non-trivial term $\sim G^2$ will then again induce a nontrivial fixed point, making quantum gravity in $2+\epsilon$ dimensions asymptotically safe.

This result exemplifies a tentative scenario in which the double-scaling limit in tensor models describes the same phase transition as a non-Gau\ss{}ian fixed point in the asymptotic safety scenario in continuum gravity: I.e., both could be different sides of the same picture, as indicated by the possibility to derive a beta-function for $G$ featuring an asymptotically safe fixed point from the double-scaling limit. 

An even more interesting scenario would be if both matter and gravitational degrees of freedom could be encoded in the dynamics of a tensor model. The existence of multicritical points in matrix models, corresponding to conformal matter coupled to gravity, shows that such a scenario works in two dimensions. To understand whether a similar scenario could work in four dimensions, one should compare the critical exponents obtained on the tensor model side, to those obtained within asymptotic safety under the coupling to matter.
On the continuum side, some critical exponents for gravity in the presence of matter degrees of freedom are known \cite{Dona:2013qba}, and also critical exponents corresponding to matter operators at the interacting fixed points are (partially) known. While currently a complete catalogue of all relevant operators at the interacting fixed point of gravity and matter is still work in progress, its availability would open the door to investigate a scenario where matter and gravity degrees of freedom are both encoded in a tensor model.

\section{Conclusions}

In this paper we advance the functional Renormalization Group as an exploratory tool for the continuum limit in matrix and tensor models for quantum gravity. 

In particular, we develop a new, self-consistent approximation which allows to obtain useful results: The use of the matrix size $N$ as a Renormalization Group scale implies a breaking of the $U(N)$ invariance of the matrix model, which is encoded in the Ward identity \Eqref{equ:WTI}. We find indications that the systematic deviation of the relevant critical exponent we found in \cite{Eichhorn:2013isa} is due to breaking of this Ward identity. The Ward identity tells us that in order to obtain a $U(N)$ symmetric continuum limit, we have to consider truncations that include a fine-tuned amount of symmetry breaking operators.

A crucial new observation in this paper is that a restriction to tadpole diagrams allows to solve the Ward identity in a self-consistent approximation with a truncation that contains only symmetric operators. Thus the effect of symmetry-breaking terms can be consistently neglected when approximating the Wetterich equation by its tadpole part. Most importantly, this approximation is well-adapted to discover the double-scaling limit in matrix and tensor models, since the fixed-point value of the couplings corresponds to the radius of convergence of the perturbative expansion of the partition function, and therefore lies at small values. At the corresponding RG fixed point, all couplings are thus much smaller than one, and loop diagrams with a higher number of vertices are therefore suppressed. 

As an added bonus, the restriction to the tadpole approximation very significantly reduces the computational complexity, in particular at the level of multiple-trace operators.

Moreover, the restriction to the tadpole approximation enables us for the first time to confirm the existence of multicritical points for matrix models within the FRG framework, both numerically and by analytic considerations. These RG fixed points  correspond to continuum limits of quantum gravity coupled to conformal matter degrees of freedom. It is particularly reassuring that, within the tadpole approximation, we find the numerical value of the leading critical exponent within about 1\% of its analytical value $4/5$ for the third and fourth multicritical point. This is a significant improvement over our previous results \cite{Eichhorn:2013isa} and shows that the FRG framework is not only a qualitative exploratory tool, but also a useful framework to obtain quantitative results with comparatively little computational effort.

We then discuss a tentative connection between the double-scaling limit in matrix and tensor models, and the asymptotic safety scenario, which provides an ultraviolet completion for continuum quantum gravity. It is conceivable that both describe the same phase transition, albeit seen from different phases: Coming from the ``pregeometric" phase, described by a matrix model, a ``condensation" of space-time building blocks occurs at a phase transition to a geometric phase. This phase can be described using continuum fields such as the metric. The phase transition is then visible as a fixed point of the Renormalization Group flow in this phase. While currently there is no solid evidence for such a scenario in four dimensions, we discuss hints for its realization in two dimensions: There, one can explicitly derive the beta-function for the Newton coupling from the double scaling limit. It contains a term $\sim G^2$ with a negative coefficient, which, in $d = 2+\epsilon$, induces an interacting fixed point for $G$. 
This is precisely the result that can be obtained explicitly by calculating the beta function in the continuum, and thus provides a hint that the double-scaling limit -- or more generally continuum limit -- of matrix models could be related to asymptotic safety.

\emph{Acknowledgements}
We acknowledge helpful discussions with R. Gurau and V. Rivasseau on tensor models. We would also like to thank J. Ambjorn for extensive discussions on the relation of the matrix model double scaling limit to the continuum beta function.
This research was supported in part by Perimeter Institute for Theoretical Physics and in part by the National Science and Research Council of Canada through a grant to the University of New Brunswick. Research at Perimeter Institute is supported by the Government of Canada
through Industry Canada and by the Province of Ontario through the Ministry of Research and Innovation.  T.K. is grateful for hospitality at the Perimeter Institute where part of the work for this paper was completed.


\begin{thebibliography}{99}
\bibitem{Eichhorn:2013isa} 
  A.~Eichhorn and T.~Koslowski,
  Phys.\ Rev.\ D {\bf 88}, 084016 (2013)
  [arXiv:1309.1690 [gr-qc]].

\bibitem{Ambjorn:1991pq} 
  J.~Ambjorn and J.~Jurkiewicz,
  Phys.\ Lett.\ B {\bf 278}, 42 (1992).

\bibitem{Ambjorn:2012jv} 
  J.~Ambjorn, A.~Goerlich, J.~Jurkiewicz and R.~Loll,
  Phys.\ Rept.\  {\bf 519}, 127 (2012)
  [arXiv:1203.3591 [hep-th]].

\bibitem{Ambjorn:2011cg} 
  J.~Ambjorn, S.~Jordan, J.~Jurkiewicz and R.~Loll,
  Phys.\ Rev.\ Lett.\  {\bf 107}, 211303 (2011)
  [arXiv:1108.3932 [hep-th]].

\bibitem{Rivasseau:2012yp} 
  V.~Rivasseau,
  arXiv:1209.5284 [hep-th].

\bibitem{Rivasseau:2011hm} 
  V.~Rivasseau,
  AIP Conf.\ Proc.\  {\bf 1444}, 18 (2011)
  [arXiv:1112.5104 [hep-th]].


%
\bibitem{Boulatov:1992vp} 
  D.~V.~Boulatov,
  Mod.\ Phys.\ Lett.\ A {\bf 7}, 1629 (1992)
  [hep-th/9202074].

\bibitem{Freidel:2005qe} 
  L.~Freidel,
  Int.\ J.\ Theor.\ Phys.\  {\bf 44}, 1769 (2005)
  [hep-th/0505016].

\bibitem{Oriti:2007qd} 
  D.~Oriti,
  PoS QG {\bf -PH}, 030 (2007)
  [arXiv:0710.3276 [gr-qc]].

\bibitem{Oriti:2011jm} 
  D.~Oriti,
  arXiv:1110.5606 [hep-th].


\bibitem{Weingarten:1982mg} 
  D.~Weingarten,
  Nucl.\ Phys.\ B {\bf 210}, 229 (1982);
  F.~David,
  Nucl.\ Phys.\ B {\bf 257}, 45 (1985);
  Nucl.\ Phys.\ B {\bf 257}, 543 (1985);
  J.~Ambjorn, B.~Durhuus and J.~Frohlich,
  Nucl.\ Phys.\ B {\bf 257}, 433 (1985);
  V.~A.~Kazakov, A.~A.~Migdal and I.~K.~Kostov,
  Phys.\ Lett.\ B {\bf 157}, 295 (1985);
  D.~V.~Boulatov, V.~A.~Kazakov, A.~A.~Migdal and I.~K.~Kostov,
  Phys.\ Lett.\ B {\bf 174}, 87 (1986);
  Nucl.\ Phys.\ B {\bf 275}, 641 (1986).





\bibitem{AlvarezGaume:1991rm} 
  L.~Alvarez-Gaume,
  Helv.\ Phys.\ Acta {\bf 64}, 359 (1991).

\bibitem{Ginsparg:1991bi} 
  P.~H.~Ginsparg,
  hep-th/9112013.

\bibitem{David:1992jw} 
  F.~David,
  hep-th/9303127.

%
\bibitem{Di Francesco:1993nw} 
  P.~Di Francesco, P.~H.~Ginsparg and J.~Zinn-Justin,
  Phys.\ Rept.\  {\bf 254}, 1 (1995)
  [hep-th/9306153].

\bibitem{Ambjorn:1994yv} 
  J.~Ambjorn,
  hep-th/9411179.

\bibitem{Douglas:1989ve} 
  M.~R.~Douglas and S.~H.~Shenker,
  Nucl.\ Phys.\ B {\bf 335}, 635 (1990).

\bibitem{Brezin:1990rb} 
  E.~Brezin and V.~A.~Kazakov,
  Phys.\ Lett.\ B {\bf 236}, 144 (1990).

\bibitem{Gross:1989vs} 
  D.~J.~Gross and A.~A.~Migdal,
  Phys.\ Rev.\ Lett.\  {\bf 64}, 127 (1990);
  Nucl.\ Phys.\ B {\bf 340}, 333 (1990).






\bibitem{Bonzom:2014oua} 
  V.~Bonzom, R.~Gurau, J.~P.~Ryan and A.~Tanasa,
  arXiv:1404.7517 [hep-th].

\bibitem{Dartois:2013sra} 
  S.~Dartois, R.~Gurau and V.~Rivasseau,
  JHEP {\bf 1309}, 088 (2013)
  [arXiv:1307.5281 [hep-th]].

\bibitem{Kaminski:2013maa} 
  W.~Kamiński, D.~Oriti and J.~P.~Ryan,
  New J.\ Phys.\  {\bf 16}, 063048 (2014)
  [arXiv:1304.6934 [hep-th]].

\bibitem{Gurau:2010ba} 
  R.~Gurau,
  Annales Henri Poincare {\bf 12}, 829 (2011)
  [arXiv:1011.2726 [gr-qc]];
  R.~Gurau and V.~Rivasseau,
  Europhys.\ Lett.\  {\bf 95}, 50004 (2011)
  [arXiv:1101.4182 [gr-qc]];
  R.~Gurau,
  Annales Henri Poincare {\bf 13}, 399 (2012)
  [arXiv:1102.5759 [gr-qc]];
  R.~Gurau,
  arXiv:1209.3252 [math-ph];
  R.~Gurau,
  Commun.\ Math.\ Phys.\  {\bf 330}, 973 (2014)
  [arXiv:1304.2666 [math-ph]];
  V.~Bonzom,
  JHEP {\bf 1306}, 062 (2013)
  [arXiv:1211.1657 [hep-th]];
  V.~Bonzom, R.~Gurau and V.~Rivasseau,
  Phys.\ Rev.\ D {\bf 85}, 084037 (2012)
  [arXiv:1202.3637 [hep-th]].

\bibitem{Geloun:2012bz} 
  J.~Ben Geloun and E.~R.~Livine,
  J.\ Math.\ Phys.\  {\bf 54}, 082303 (2013)
  [arXiv:1207.0416 [hep-th]];
  J.~Ben Geloun,
  Class.\ Quant.\ Grav.\  {\bf 29}, 235011 (2012)
  [arXiv:1205.5513 [hep-th]];
  J.~Ben Geloun and D.~O.~Samary,
  Annales Henri Poincare {\bf 14}, 1599 (2013)
  [arXiv:1201.0176 [hep-th]];
  J.~Ben Geloun and V.~Rivasseau,
  Commun.\ Math.\ Phys.\  {\bf 318}, 69 (2013)
  [arXiv:1111.4997 [hep-th]].

\bibitem{Oriti:2013jga} 
  D.~Oriti,
  Stud.\ Hist.\ Philos.\ Mod.\ Phys.\  {\bf 46}, 186 (2014)
  [arXiv:1302.2849 [physics.hist-ph]].

\bibitem{Brezin:1992yc} 
  E.~Brezin and J.~Zinn-Justin,
  Phys.\ Lett.\ B {\bf 288}, 54 (1992)
  [hep-th/9206035].

%
\bibitem{Ayala:1993fj} 
  C.~Ayala,
  Phys.\ Lett.\ B {\bf 311}, 55 (1993)
  [hep-th/9304090].

\bibitem{Higuchi:1993np} 
  S.~Higuchi, C.~Itoi and N.~Sakai,
  Phys.\ Lett.\ B {\bf 312}, 88 (1993)
  [hep-th/9303090];
  S.~Higuchi, C.~Itoi and N.~Sakai,
  Prog.\ Theor.\ Phys.\ Suppl.\  {\bf 114}, 53 (1993)
  [hep-th/9307154];
  S.~Higuchi, C.~Itoi, S.~Nishigaki and N.~Sakai,
  Phys.\ Lett.\ B {\bf 318}, 63 (1993)
  [hep-th/9307116].

\bibitem{Bonnet:1998ei} 
  G.~Bonnet and F.~David,
  Nucl.\ Phys.\ B {\bf 552}, 511 (1999)
  [hep-th/9811216].

\bibitem{Abbott:1980hw}
  L.~F.~Abbott,
  Nucl.\ Phys.\  B {\bf 185}, 189 (1981).

%
\bibitem{Reuter:1996cp} 
  M.~Reuter,
  Phys.\ Rev.\ D {\bf 57}, 971 (1998)
  [hep-th/9605030].

\bibitem{Donkin:2012ud} 
  I.~Donkin and J.~M.~Pawlowski,
  arXiv:1203.4207 [hep-th].

\bibitem{Ambjorn:2014gsa} 
  J.~Ambjorn, A.~Görlich, J.~Jurkiewicz, A.~Kreienbuehl and R.~Loll,
  Class.\ Quant.\ Grav.\  {\bf 31}, 165003 (2014)
  [arXiv:1405.4585 [hep-th]].

\bibitem{Cooperman:2014owa} 
  J.~H.~Cooperman,
  arXiv:1406.4531 [gr-qc].

\bibitem{Dona:2013qba} 
  P.~Donà, A.~Eichhorn and R.~Percacci,
  arXiv:1311.2898 [hep-th].

\bibitem{Kazakov:1989bc} 
  V.~A.~Kazakov,
  Mod.\ Phys.\ Lett.\ A {\bf 4}, 2125 (1989).

\bibitem{Staudacher:1989fy} 
  M.~Staudacher,
  Nucl.\ Phys.\ B {\bf 336}, 349 (1990).


\bibitem{Ambjorn:1992gw} 
  J.~Ambjorn, L.~Chekhov, C.~F.~Kristjansen and Y.~Makeenko,
  Nucl.\ Phys.\ B {\bf 404}, 127 (1993)
  [Erratum-ibid.\ B {\bf 449}, 681 (1995)]
  [hep-th/9302014].



\bibitem{Wetterich:1993yh}
C.~Wetterich,
Phys.\ Lett.\ B {\bf 301}, 90 (1993).



\bibitem{Berges:2000ew}
  J.~Berges, N.~Tetradis and C.~Wetterich,
  %
  Phys.\ Rept.\  {\bf 363} (2002) 223
  [hep-ph/0005122].
%
  J.~Polonyi,
  Central Eur.\ J.\ Phys.\  {\bf 1} (2003) 1. 
 [hep-th/0110026];
%
  J.~M.~Pawlowski,
  Annals Phys.\  {\bf 322} (2007) 2831 
  [arXiv:hep-th/0512261];
%
  B.~Delamotte,
  Lect.\ Notes Phys.\  {\bf 852}, 49 (2012)
  [cond-mat/0702365 [COND-MAT]];
  O.~J.~Rosten,
  arXiv:1003.1366 [hep-th].

\bibitem{Gies:2006wv}
  H.~Gies, in {\it Renormalization Group and Effective Field Theory Approaches to Many-Body Systems,
Lecture Notes in Physics, Vol 852}, pp 287-348 (2012)
  arXiv:hep-ph/0611146; 

%
\bibitem{Litim:2001up} 
  D.~F.~Litim,
  Phys.\ Rev.\ D {\bf 64}, 105007 (2001)
  [hep-th/0103195].

\bibitem{Litim:2000ci} 
  D.~F.~Litim,
  Phys.\ Lett.\ B {\bf 486}, 92 (2000)
  [hep-th/0005245].

\bibitem{Falls:2013bv} 
  K.~Falls, D.~F.~Litim, K.~Nikolakopoulos and C.~Rahmede,
  arXiv:1301.4191 [hep-th].


\bibitem{Sfondrini:2010zm}
  A.~Sfondrini and T.~A.~Koslowski,
  Int.\ J.\ Mod.\ Phys.\ A {\bf 26} (2011) 4009
  [arXiv:1006.5145 [hep-th]].

\bibitem{Weinberg:1976xy} 
  S.~Weinberg,
  HUTP-76-160.

\bibitem{Weinberg:1980gg}
  S.~Weinberg,
{\it  In *Hawking, S.W., Israel, W.: General Relativity*, 790-831}
(Cambridge University Press, Cambridge, 1980).

\bibitem{AS_reviews}
  M.~Niedermaier and M.~Reuter,
  Living Rev.\ Rel.\  {\bf 9}, 5 (2006);
  M.~Niedermaier,
  Class.\ Quant.\ Grav.\  {\bf 24}, R171 (2007)
  [gr-qc/0610018];
  R.~Percacci,
  In Oriti, D. (ed.): ``Approaches to quantum gravity'' 111-128
  [arXiv:0709.3851 [hep-th]];
  D.~F.~Litim,
  arXiv:0810.3675 [hep-th];
  Phil.\ Trans.\ Roy.\ Soc.\ Lond.\ A {\bf 369}, 2759 (2011)
  [arXiv:1102.4624 [hep-th]];
  R.~Percacci,
  arXiv:1110.6389 [hep-th];
  M.~Reuter and F.~Saueressig,
  New J.\ Phys.\  {\bf 14}, 055022 (2012)
  [arXiv:1202.2274 [hep-th]];
  arXiv:1205.5431 [hep-th];
%
  S.~Nagy,
  arXiv:1211.4151 [hep-th].


\bibitem{Rivasseau:2014ima} 
  V.~Rivasseau,
  arXiv:1407.0284 [hep-th].

\bibitem{Percacci:2010af} 
  R.~Percacci and G.~P.~Vacca,
  Class.\ Quant.\ Grav.\  {\bf 27}, 245026 (2010)
  [arXiv:1008.3621 [hep-th]].

\bibitem{Ambjorn:2011cg} 
  J.~Ambjorn, S.~Jordan, J.~Jurkiewicz and R.~Loll,
  Phys.\ Rev.\ Lett.\  {\bf 107}, 211303 (2011)
  [arXiv:1108.3932 [hep-th]];
  J.~Ambjorn, S.~Jordan, J.~Jurkiewicz and R.~Loll,
  Phys.\ Rev.\ D {\bf 85}, 124044 (2012)
  [arXiv:1205.1229 [hep-th]].


\bibitem{Reuter:2008qx} 
  M.~Reuter and H.~Weyer,
  Phys.\ Rev.\ D {\bf 80}, 025001 (2009)
  [arXiv:0804.1475 [hep-th]].



\bibitem{Ambjorn:2005db} 
  J.~Ambjorn, J.~Jurkiewicz and R.~Loll,
  Phys.\ Rev.\ Lett.\  {\bf 95}, 171301 (2005)
  [hep-th/0505113];
  A.~Gorlich,
  arXiv:1111.6938 [hep-th];
  D.~Benedetti and J.~Henson,
  Phys.\ Rev.\ D {\bf 80}, 124036 (2009)
  [arXiv:0911.0401 [hep-th]];
  C.~Anderson, S.~J.~Carlip, J.~H.~Cooperman, P.~Horava, R.~K.~Kommu and P.~R.~Zulkowski,
  Phys.\ Rev.\ D {\bf 85}, 044027 (2012)
  [arXiv:1111.6634 [hep-th]].

\bibitem{Lauscher:2005qz} 
  O.~Lauscher and M.~Reuter,
  JHEP {\bf 0510}, 050 (2005)
  [hep-th/0508202];
  O.~Lauscher and M.~Reuter,
  In *Fauser, B. (ed.) et al.: Quantum gravity* 293-313
  [hep-th/0511260];
  M.~Reuter and F.~Saueressig,
  JHEP {\bf 1112}, 012 (2011)
  [arXiv:1110.5224 [hep-th]];
  S.~Rechenberger and F.~Saueressig,
  Phys.\ Rev.\ D {\bf 86}, 024018 (2012)
  [arXiv:1206.0657 [hep-th]];
  G.~Calcagni, A.~Eichhorn and F.~Saueressig,
  Phys.\  Rev.\ D {\bf 87}, 124028 (2013)
  [arXiv:1304.7247 [hep-th]].

\bibitem{Horava:2009if} 
  P.~Horava,
  Phys.\ Rev.\ Lett.\  {\bf 102}, 161301 (2009)
  [arXiv:0902.3657 [hep-th]];
  T.~P.~Sotiriou, M.~Visser and S.~Weinfurtner,
  Phys.\ Rev.\ Lett.\  {\bf 107}, 131303 (2011)
  [arXiv:1105.5646 [gr-qc]].


\bibitem{twoplusepsilon} 
  S.~M.~Christensen and M.~J.~Duff,
  Phys.\ Lett.\ B {\bf 79}, 213 (1978);
  R.~Gastmans, R.~Kallosh and C.~Truffin,
  Nucl.\ Phys.\ B {\bf 133}, 417 (1978);
  H.~Kawai and M.~Ninomiya,
  Nucl.\ Phys.\ B {\bf 336}, 115 (1990);
  I.~Jack and D.~R.~T.~Jones,
  Nucl.\ Phys.\ B {\bf 358}, 695 (1991);
 H.~Kawai, Y.~Kitazawa and M.~Ninomiya,
  Nucl.\ Phys.\ B {\bf 393}, 280 (1993)
  [hep-th/9206081];
  Nucl.\ Phys.\ B {\bf 404}, 684 (1993)
  [hep-th/9303123];
  M.~Fukuma, N.~Ishibashi, H.~Kawai and M.~Ninomiya,
  Nucl.\ Phys.\ B {\bf 427}, 139 (1994)
  [hep-th/9312175];
 H.~Kawai, Y.~Kitazawa and M.~Ninomiya,
  Prog.\ Theor.\ Phys.\ Suppl.\  {\bf 114}, 149 (1993);
  T.~Aida, Y.~Kitazawa, H.~Kawai and M.~Ninomiya,
  Nucl.\ Phys.\ B {\bf 427}, 158 (1994)
  [hep-th/9404171];
  J.~Nishimura, S.~Tamura and A.~Tsuchiya,
  Mod.\ Phys.\ Lett.\ A {\bf 9}, 3565 (1994)
  [hep-th/9405059];
  H.~Kawai, Y.~Kitazawa and M.~Ninomiya,
  Nucl.\ Phys.\ B {\bf 467}, 313 (1996)
  [hep-th/9511217];
  T.~Aida and Y.~Kitazawa,
  Nucl.\ Phys.\ B {\bf 491}, 427 (1997)
  [hep-th/9609077].










\end{thebibliography}
\end{document}